\documentclass[iop]{emulateapj}
\usepackage{psfig}

\usepackage{amsmath}
\usepackage{textcomp}
\usepackage[colorlinks=true,citecolor=blue]{hyperref}
\usepackage{graphicx,color}
\usepackage{dcolumn}
\usepackage{times}

\shorttitle{The Nature of $\gamma$-NLSy1 Galaxies}
\shortauthors{V S Paliya et al.}

\begin{document}


\title{The Nature of $\gamma$-ray Loud Narrow Line Seyfert I Galaxies PKS 1502+036 and PKS 2004$-$447}


\author{Vaidehi S. Paliya\altaffilmark{1,2}, C. S. Stalin\altaffilmark{1}, 
Amit Shukla\altaffilmark{1}, S. Sahayanathan\altaffilmark{3}}
\altaffiltext{1}{Indian Institute of Astrophysics, Block II, Koramangala, 
Bangalore-560034, India}
\altaffiltext{2}{School of Inter-Disciplinary and Trans-Disciplinary Studies, 
IGNOU, New Delhi-110068, India}
\altaffiltext{3}{Astrophysical Science Division, Bhabha Atomic Research Center,
Mumbai 400085, India}
\email{vaidehi@iiap.res.in}





\begin{abstract}
Variable $\gamma$-ray emission has been discovered in five Radio-loud 
Narrow Line Seyfert 1 (NLSy1) galaxies by the Large Area Telescope (LAT) 
onboard the {\it Fermi} Gamma-Ray Space Telescope. This has clearly 
demonstrated that these NLSy1 galaxies do have relativistic jets similar to 
two other cases of $\gamma$-ray emitting Active Galactic Nuclei (AGN), namely 
blazars and radio galaxies. We present here our results on the multi-band 
analysis of two $\gamma$-ray emitting NLSy1 galaxies namely 
PKS 1502+036 ($z$ = 0.409) and PKS 2004$-$447 ($z$ = 0.240) using archival 
data. We generate multi-band long term light curves of these sources, 
 build their spectral energy distribution (SED) and model 
them using 
an one zone leptonic model. They resemble more to the
SEDs of the flat spectrum radio quasar (FSRQ) class of AGN. We then compare the SEDs of these two sources with two other {\it Fermi} detected AGN along 
the traditional blazar sequence, namely the BL Lac Mrk 421 ($z$ = 0.03) and the FSRQ 3C 454.3 ($z$ = 0.86). The SEDs of both PKS 1502+036 and PKS 2004$-$447 are found to be intermediate to the SEDs of Mrk 421 and 3C 454.3. In the $\gamma$-ray spectral index v/s $\gamma$-ray luminosity plane, both these NLSy1 galaxies occupy a distinct position, wherein, they have luminosity between Mrk 421 and 3C 454.3, however steep $\gamma$-ray spectra similar to 3C 454.3. Their Compton dominance as well as their X-ray spectral slope also lie between Mrk 421 and 3C 454.3. We argue that the physical properties of both PKS 1502+036 and PKS 2004$-$447 are in general similar to blazars and intermediate between FSRQs and BL Lac objects and these sources thus could fit into the traditional blazar sequence.

\end{abstract}


\keywords{galaxies: active $-$ gamma rays: galaxies $-$ quasars: 
individual(PKS 1502+036,
PKS 2004-447 $-$ galaxies: Seyfert}



\section{Introduction}{\label{sec1}}
The extra-galactic sky in the high energy $\gamma$-ray band is dominated by a special class of Active Galactic Nuclei (AGN) called blazars (\citealt{2010ApJ...715..429A}). These blazars which come under two sub-classes namely Flat Spectrum Radio Quasars (FSRQs) and BL Lacs (see, for e.g., \citealt{1995PASP..107..803U}), have similar broad band continuum properties, such that they emit variable non-thermal emission over the entire electromagnetic
spectrum. They both have flat radio spectra ($\alpha_r < 0.5; S_{\nu} \propto \nu^{-\alpha}$), at GHz frequencies, high optical polarization, exhibits superluminal pattern at radio-frequencies and also show rapid flux and polarization variations (\citealt{1995ARA&A..33..163W,2005A&A...442...97A}). However, the distinction between FSRQs and BL Lacs is made on the difference in the strength of their optical emission lines. BL Lacs have weak optical 
emission lines with rest frame equivalent widths less than 5 \AA (\citealt{1991ApJS...76..813S}; \citealt{1991ApJ...374..431S}). The observed broad band 
emission of blazars is believed to originate from relativistic jets which are
aligned closely to the observer (\citealt{1984RvMP...56..255B}). Their spectral energy distribution (SED) shows two prominent and distinct peaks: the low 
energy peak due to synchrotron emission, lies between the far-infrared and 
the X-ray band, and the high energy peak due to inverse Compton (IC) 
radiation, lies in the MeV to GeV energy range. The seed photons for IC emission can be synchrotron photons itself (Synchrotron Self Compton, SSC;  \citealt{1981ApJ...243..700K,1985ApJ...298..114M,1989ApJ...340..181G}) or photons external to the jet 
(external Compton, EC; \citealt{1987ApJ...322..650B,1989ApJ...340..162M,1992A&A...256L..27D}). The plausible sources of seed photons for EC can be accretion 
disc (\citealt{1993ApJ...416..458D,1997A&A...324..395B}), 
broad line region (BLR) 
(\citealt{1994ApJ...421..153S,1996MNRAS.280...67G}), 
dusty torus (\citealt{2000ApJ...545..107B,2008MNRAS.387.1669G}), 
bulge of the host galaxy, or cosmic microwave background 
radiation (\citealt{2009MNRAS.397..985G}).  
In the $\gamma$-ray regime, FSRQs are more luminous and exhibit 
softer spectral index than BL Lacs (\citealt{2009ApJ...700..597A}). It has been noted by \citet{1998MNRAS.299..433F} and  \citet{1998MNRAS.301..451G} that both FSRQs and BL Lacs follow the so called blazar sequence. However, recently \citet{2012MNRAS.420.2899G} has argued that the existence of blazar sequence could 
be due to selection effects. 

In addition to blazars which are dominating the $\gamma$-ray 
extra-galactic sky, 
{\it Fermi} has also discovered variable $\gamma$-ray emission from 
five Narrow Line Seyfert 1 (NLSy1) 
galaxies \citep{2009ApJ...699..976A,2009ApJ...707L.142A,
2012MNRAS.426..317D,2011MNRAS.413.2365C}. A few more NLSy1 sources are
suspected to be $\gamma$-ray emitters, although the detection significance
is still low \citep{2011nlsg.confE..24F}. NLSy1 galaxies are a separate class of AGN with peculiar properties. Their optical spectra are similar to the conventional broad line Seyfert 1 galaxies except that they have narrow Balmer lines 
(FWHM (H$_{\beta}$) $<$ 2000 km/sec), weak [O~{\sc iii}] 
([O~{\sc iii}]/H$_{\beta} <$ 3) and strong optical Fe~{\sc ii} 
lines \citep{1985ApJ...297..166O, 1989ApJ...342..224G}. They also have steep soft X-ray spectra \citep{1996A&A...305...53B, 1996A&A...309...81W, 1999ApJS..125..297L} and show rapid X-ray flux variations \citep{1995MNRAS.277L...5P,1999ApJS..125..317L}. These peculiar observational characteristics are attributed to them having low-mass black holes ($\sim$10$^{6}$$-$10$^{8}$ M$_{\odot}$) and accreting close to the Eddington limit \citep{2000ApJ...542..161P, 2000NewAR..44..419H,2004ApJ...606L..41G,2006ApJS..166..128Z, 2012AJ....143...83X}.
However, recently, \citet{2012arXiv1212.1181C}, using multi-wavelength data, 
have shown that radio-loud NLSy1 (RL-NLSy1) galaxies have black 
hole masses similar to blazars. These  RL-NLSy1 galaxies comprising about 7\% of NLSy1 galaxy population, exhibit compact core-jet structure, flat/inverted radio spectra, high brightness temperature and superluminal motion \citep{2006AJ....132..531K,2006PASJ...58..829D}.  Recently, kiloparsec-scale radio 
structures are found in six RL-NLSy1 galaxies (\citealt{2012ApJ...760...41D}). 
The SEDs of some RL-NLSy1 galaxies are found to be similar to those of high 
frequency peaked BL Lac objects and some of them also show 
a hard X-ray component in their bright optical/UV state 
(\citealt{2008ApJ...685..801Y}). Though these characteristics are indicative of the presence of relativistic jets in these sources that are closely aligned to the observer, the detection of $\gamma$-rays in five of them has confirmed their presence \citep{2009ApJ...699..976A,2009ApJ...707L.142A,2012MNRAS.426..317D} similar to the blazar class of AGN. Also, it has been recently reported that the optical and infra-red (IR) flux variations of some of these sources are similar to blazars (\citealt{2010ApJ...715L.113L,2013MNRAS.428.2450P,2012ApJ...759L..31J,2013ApJ...762..124M}). However, these similarities are at odds with the 
common belief that NLSy1 galaxies are hosted in spiral galaxies with 
low mass black holes,  while, blazars with powerful relativistic
jets are hosted in elliptical galaxies 
with high mass black holes. Therefore, detailed studies of these sources will enable us to understand the physical properties of relativistic jets hosted in AGN with low mass central black holes and high accretion rates. In this work, we have carried out a detailed analysis of two such sources, namely, PKS 1502+036 and PKS 2004$-$447 to see (i) the similarities and/or differences of them vis-a-vis the blazars detected by {\it Fermi} and (ii) their place in the traditional blazar sequence. The detailed analysis of these sources are 
presented here for the first time. Throughout the work, we adopt $\Omega_{m}$=0.27, $\Omega_{\varLambda}$=0.73 and  Hubble constant {\it H}$_{0}$ = 70 km s$^{-1}$ Mpc$^{-1}$.


\section{Sample}{\label{sec2}}
Among the 5 known $\gamma$-ray emitting NLSy1 ($\gamma$-NLSy1) galaxies, we 
have selected two of them, namely, PKS 1502+036 and PKS 2004$-$447. 
They have intermediate redshifts when compared to other $\gamma$-NLSy1 galaxies. PKS 1502+036 has a very compact core on observations with the Very 
Large Array (VLA; \citealt{2012arXiv1205.0402O, 2011AJ....141..182K}). However, when observed at 15 GHz, using Very Long Baseline Array (VLBA) at parsec scale resolution, it was found to have a core-jet structure (\citealt{2012arXiv1205.0402O}). This NLSy1 galaxy, with a central 
black hole mass of $\sim$ 4 $\times $10$^{6 }$ M$_{\odot}$, has a brightness temperature of $\sim$10$^{12}$ K \citep{2008ApJ...685..801Y}. It is a faint, but persistent $\gamma$-ray emitter (\citealt{2011MNRAS.413.2365C,2009ApJ...707L.142A}) and also show intra-night optical variability (INOV; \citealt{2013MNRAS.428.2450P}) and rapid IR variability (\citealt{2012ApJ...759L..31J}).

PKS 2004$-$447 has been classified as a NLSy1 galaxy by \citet{2001ApJ...558..578O} and \citet{2006MNRAS.370..245G} but \citet{2006AJ....132..531K} classify  it as narrow-line radio galaxy owing to the presence of weak Fe II lines in its optical spectra. It has a steep spectrum above 8.4 GHz \citep{2012arXiv1205.0402O}, however, has a flat spectrum below 5 GHz \citep{2007ApJS..171...61H}. At 20 GHz, it shows a fractional polarization of 3.8$\%$ \citep{2010MNRAS.402.2403M}. The general properties of both sources are given in Table \ref{tab1}.

\begin{table*}
\begin{center}
{
\small
\caption{List of the $\gamma$-NLSy1 galaxies  monitored in this work and the results of the analysis of about four years of {\it Fermi} LAT data. Column 
informations are as follows: (1) Name; (2) right ascension; (3) declination; (4) redshift; (5) absolute B magnitude; (6) apparent V magnitude; (7) radio spectral index; (8) radio loudness parameter;  (9) 0.1$-$300 GeV flux in the units of 10$^{-8}$ ph cm$^{-2}$ s$^{-1}$; (10) photon index obtained from {\it Fermi}-LAT data analysis; (11) TS and (12) $N_H$ \label{tab1}}
\begin{tabular}{lcccccrrcccc}
\tableline\tableline
Name  & RA (2000) & Dec (2000) & z\tablenotemark{a} & M$_{B}$\tablenotemark{a} & V\tablenotemark{a} & $\alpha_R$\tablenotemark{b} & R\tablenotemark{c}  & F$_{0.1 - 300 GeV}$ &  $\Gamma$ & TS & $N_H$\tablenotemark{d}\\
      &  (h m s)  & (d m s)    &                    & (mag)                    & (mag)              &                             &                     &                    &           &  & \\
(1)   & (2)       & (3)        & (4)                &  (5)                     &  (6)               &  (7)                        & (8)                 &  (9)               &   (10)    & (11) & (12)  \\
\tableline
PKS 1502+036   & 15:05:06.5 & +03:26:31 & 0.409 & $-$22.8 & 18.64 & 0.41 & 3364 & 5.15 $\pm$ 0.42 & 2.67 $\pm$ 0.06 & 419.28 & 3.93 \\
PKS 2004$-$447 & 20:07:55.1 & -44:34:43 & 0.240 & $-$21.6 & 19.30 &0.38  & 6358 & 1.66 $\pm$ 0.33 & 2.58 $\pm$ 0.12 &  67.67 & 3.17 \\ \hline

\tableline
\end{tabular}
\tablenotetext{1}{\citet{2010A&A...518A..10V}}
\tablenotetext{2}{Radio spectral index calculated using 6 cm and 20 cm flux densities given by \citet{2010A&A...518A..10V}}
\tablenotetext{3}{\citet{2011nlsg.confE..24F}}
\tablenotetext{4}{Galactic absorption in units of 10$^{20}$ cm$^{-2}$ from \citet{2005A&A...440..775K}}
}
\end{center}
\end{table*}

\section{Data Reduction and Analysis}{\label{sec3}}
\subsection{{\it FERMI}-Large Area Telescope}{\label{subsec1}}

The {\it Fermi}-Large Area Telescope (LAT; \citealt{2009ApJ...697.1071A}) data 
used in this work were collected over the last four years of {\it Fermi} operation, from 2008 August 05 00:00:00 UT to 2012 July 24 00:00:00 UT. LAT photons of event class 2 between 100 MeV to 300 GeV were extracted from {\it Fermi} Science Support Center\footnote[4]{http://fermi.gsfc.nasa.gov/ssc/data} within 30$^{\circ}$ circular region of interest centered around the position of the objects. Analysis of the data was done with the LAT software package {\tt 
ScienceTools v9r27p1} along with the use of post-launch instrument response functions P7SOURCE$\_$V6 and the corresponding Galactic and isotropic diffuse background models. Additionally, a cut on the zenith angle ($<$ 100$^{\circ}$) was applied in order to exclude the contamination from Earth's-albedo. The spectral analysis were performed using an unbinned maximum likelihood method. The data were fitted with a simple power law model. In the analysis, we consider the source to be detected if test statistic (TS) $>$ 9. This corresponds to $\sim$3$\sigma$ detection \citep{1996ApJ...461..396M}. For TS $<$ 9, 2$\sigma$ upper limits were calculated.  Systematic uncertainties in the flux were estimated as 10$\%$ at 0.1 GeV, 5$\%$ at 560 MeV and 20$\%$ at 10 GeV and above\footnote[5]{http://fermi.gsfc.nasa.gov/ssc/data/analysis/LAT$\_$caveats.html}. 
Results of the average analysis are given in Table \ref{tab1}. 

\subsection{SWIFT ({\it BAT, XRT, UVOT})}{\label{subsec2}}
We have searched for the data from all the three telescopes onboard {\it Swift} (\citealt{2004ApJ...611.1005G}), namely the Burst Alert Telescope (BAT; \citealt{2005SSRv..120..143B}) operating in the 15$-$150 keV band, the X-ray telescope (XRT; \citealt{2005SSRv..120..165B}) operating in the 0.3$-$10 keV band and the Ultra-Violet Optical Telescope (UVOT; \citealt{2005SSRv..120...95R}) which can observe in 6 filters; namely {\it V, B, U, UVW1, UVM2 and UVW2}. We extracted all publicly available optical/UV/X-ray data for the sources from {\it HEASARC} archives. The data were analyzed using latest {\it Swift} tools included in {\tt HEASOFT v.6.12.0} together with calibration data files (CALDB) updated on 2012 April 02. 

Both $\gamma$-NLSy1s could not be detected in the 15$-$150 keV band. We analyzed the  XRT data using the XRTDAS software package developed at the ASI Science Data Center (ASDC) and distributed by HEASARC within the HEASoft package (v.6.12.0). Data were cleaned with standard filtering criteria and calibrated using the {\tt xrtpipeline} task with default parameters and single-to-quadruple events (grades 1$-$12). We did not find any evidence of pile-up. Events for spectral analysis were selected within a circle of 50$^{\prime\prime}$ radius, centered on the source position, while the background was selected from a nearby source-free region with 
100$^{\prime\prime}$ radius. To generate lightcurves we first rebinned the spectrum to have at least 1 count per bin (using {\tt grppha}) and then calculated the likelihood using C-statistics (\citealt{1979ApJ...228..939C}). The data were fitted with absorbed power-law model with Galactic absorption values taken from \citet{2005A&A...440..775K}. For lightcurve analysis, we have not corrected the XRT data for galactic absorption. For SED analysis, the individual XRT event files were combined together using the {\tt XSELECT} package and the average spectrum was extracted from summed event files. Depending on the total
detected counts, we grouped the spectrum so as to have at least 5 counts per bin for PKS 1502+036 and 15 counts per bin for PKS 2004$-$447. Model fitting was done using {\tt XSPEC} (\citealt{1996ASPC..101...17A}). Finally, the SED data were corrected for Galactic absorption and then binned appropriately for both sources. 


{\it Swift}-UVOT has observed these sources in all six filters for most of the observations but not always. The UVOT data were integrated with {\tt uvotimsum} task and then analyzed with {\tt uvotsource} task. The source region was chosen as a circle of 5$^{\prime\prime}$ radius for optical filters and 10$^{\prime\prime}$ radius for UV, while the 1$^{\prime}$ sized background region was extracted from nearby source-free regions. The observed magnitudes were de-reddened using the extinction laws of \citet{1989ApJ...345..245C} and converted to flux units using the zero points and conversion factors of the {\it Swift}-CALDB (\citealt{2008MNRAS.383..627P}).

\section {Results}{\label{sec4}}
\subsection{Multi-band temporal behaviour}{\label{subsec3}}

The long-term multi-band light curves of the two sources PKS 1502+036 and PKS 2004$-$447 are shown in Figure \ref{fig1} and \ref{fig2} respectively. In the $\gamma$-ray band, flux values were obtained over 30 day binning, whereas in the other bands, each flux point corresponds to one observation of {\it Swift}. For PKS 1502+036, only four sets of {\it Swift} data were available. As the 
data are sparse, no strong claims of variability could be made. However, visual examination show some hints of variability in the UVOT bands. In the $\gamma$-ray band, the source was mostly in the low activity state, however, was found to be in a relatively bright state between MJD 55592 and 55622. It is also found to be variable at 15 GHz when observed from Ovens Valley Radio Observatory (OVRO)\footnote[6]{http://www.astro.caltech.edu/ovroblazars/data/data.php}.

For PKS 2004$-$447, the sparseness of the data precluded us to provide any estimates of the variability of the source. Visual examination of the lightcurves in the UVOT bands give hints that the source has shown variability. Source has also shown significant variability in the X-ray band. When observed first time on MJD 55696, its observed flux was (4.42$\pm$1.23) $\times$ 10$^{-13}$ erg cm$^{-2}$ s$^{-1}$, which increased by $\sim$3 times when observed again on MJD 55821 but diminished to (5.36$\pm$2.20) $\times$ 10$^{-13}$ erg cm$^{-2}$ s$^{-1}$ when observed on MJD 56130. In the $\gamma$-ray band, the source was detected only in two epochs when analyzed using a 30 day binning. Comparing the counts at those two epochs we find that the source has varied in flux.

\subsection{X-ray spectral fit}{\label{subsec4}}
For PKS1502+036, all {\it Swift}/XRT observations were integrated (global exposure = 19.25 ksec) and then fitted with an absorbed power law. We find photon index, $\Gamma$ = 1.70$\pm$0.20, the normalization at 1 KeV = (4.49$\pm$0.67) $\times$ 10$^{-5}$ ph cm$^{-2}$ s$^{-1}$ keV$^{-1}$ and the integrated flux in the 0.3$-$10 keV energy band = (2.90$\pm$0.69) $\times$ 10$^{-13}$ erg cm$^{-2}$ s$^{-1}$. 

On fitting an absorbed power law model to the integrated data (global exposure = 45.89 ksec) of PKS 2004$-$447, we find $\Gamma$ = 1.53$\pm$0.09 and the normalization at 1 keV = (8.14$\pm$0.57) $\times$ 10$^{-5}$ ph cm$^{-2}$ s$^{-1}$ keV$^{-1}$. The flux in the 0.3$-$10 keV energy band = (6.26$\pm$0.73) $\times$ 10$^{-13}$ erg cm$^{-2}$ s$^{-1}$. 

\subsection{$\gamma$-ray spectral variability analysis}{\label{subsec5}}
In Figure \ref{fig3}, we show the correlation of the 0.1$-$300 GeV flux against the $\gamma$-ray spectral slope for the source PKS 1502+036, when 30 day time 
binning was used. The photon index values were obtained by fitting a power law model to the {\it Fermi}-LAT data. Clear evidence of spectral softening when brightening was found for this source when the data was fit using both the unweighted least squares method and weighted linear least squares fit that takes into account the errors on both the dependent and independent variables (\citealt{1992nrca.book.....P}). In the case of PKS 2004$-$447, we have only two detections at the epochs MJD 55232 and MJD 55802. The 
flux values at these two epochs are (7.93 $\pm$ 2.74) $\times$ 10$^{-8}$ 
and  (4.72 $\pm$ 1.91) $\times$ 10$^{-8}$  
ph cm$^{-2}$ s$^{-1}$ respectively. The photon indices at MJD 55232 and MJD 55802 are 3.12 $\pm$ 0.38 and 
2.56 $\pm$ 0.31. Thus, between the two epochs of detection, the $\gamma$-ray spectral index shows a softening when brightening trend. This behavior of the photon index is contradictory to the harder when brighter trend seen in the brightest FSRQ 3C 454.3 (\citealt{2010ApJ...710.1271A}).

\subsection{Spectral Energy Distribution}
\begin{figure*}
\begin{center}
\includegraphics[height=14cm,width=17cm]{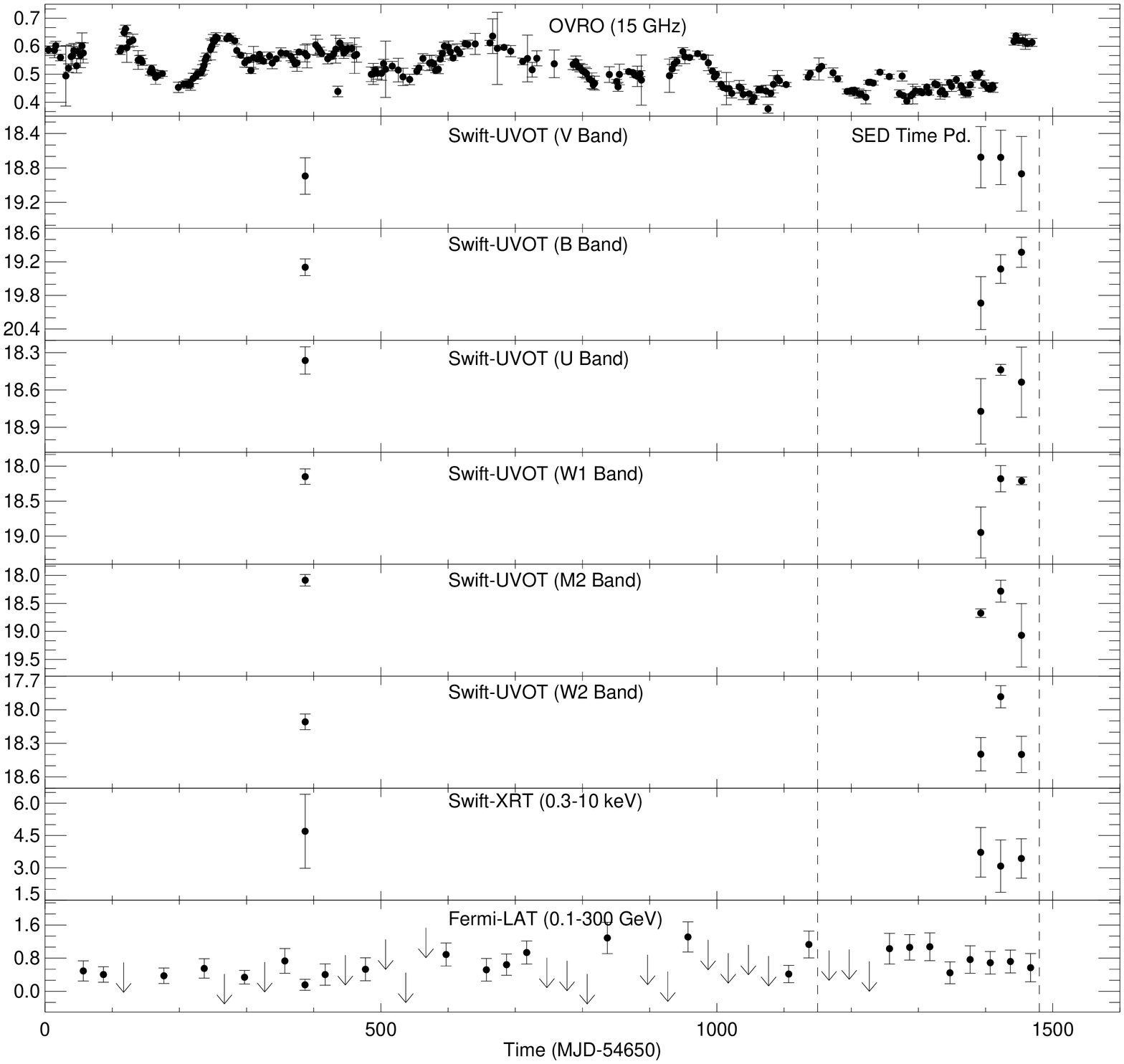}
\end{center}
\caption{Multi-band lightcurves of PKS 1502+036. The OVRO data are in Jansky. 
UVOT data are in magnitude. XRT fluxes are in units of 10$^{-13}$ erg cm$^{-2}$ s$^{-1}$, while LAT $\gamma$-ray data has the unit of 10$^{-7}$ ph cm$^{-2}$ s$^{-1}$. Here, 95\% upper limits are shown by vertical arrows. 
Dashed vertical line shows the time interval covered for SED modeling.}
\label{fig1}
\end{figure*}

\begin{figure*}
\begin{center}
\includegraphics[height=12cm,width=17cm]{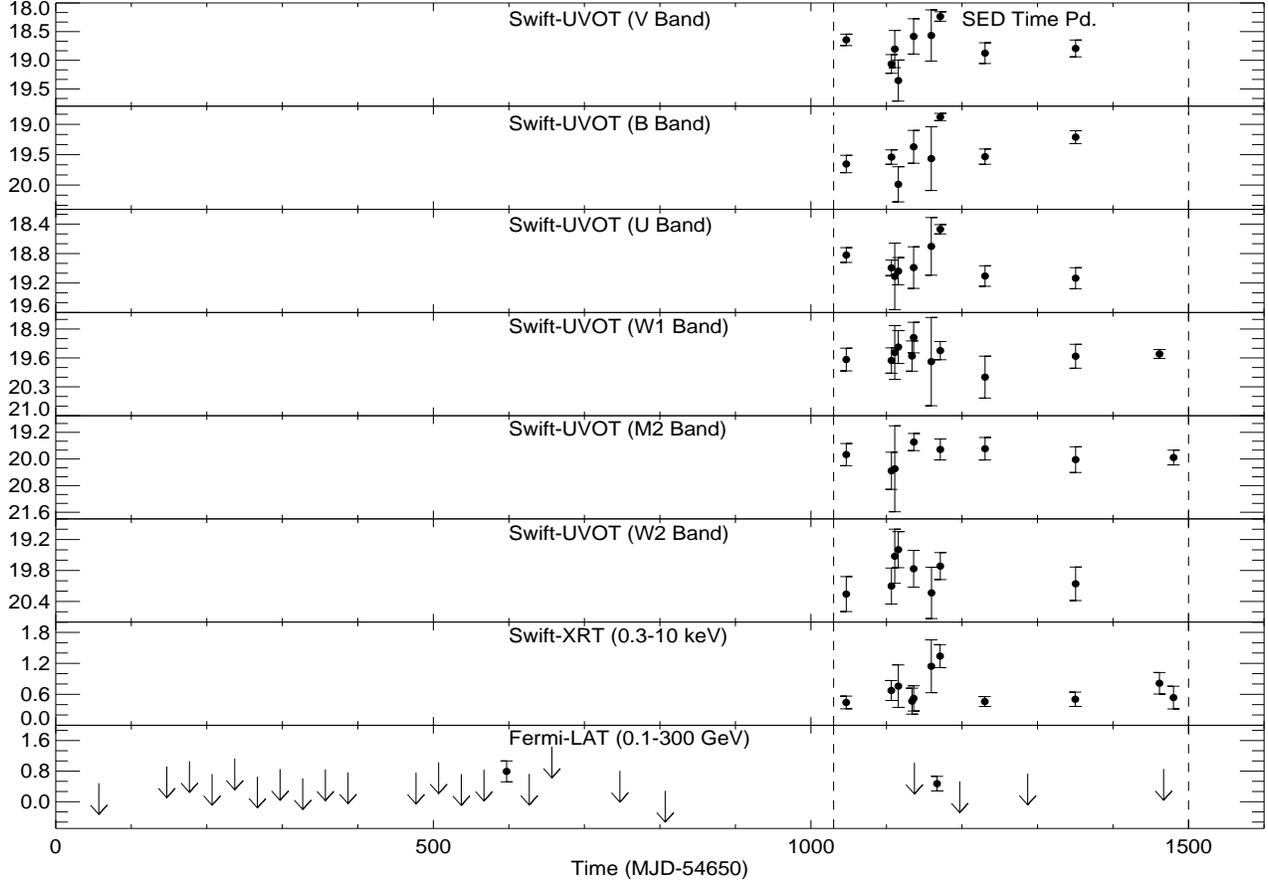}
\end{center}
\caption{Multi-band lightcurves of PKS 2004$-$447. UVOT data are in magnitude. XRT fluxes are in units of 10$^{-12}$ erg cm$^{-2}$ s$^{-1}$, while LAT $\gamma$-ray data has the unit of 10$^{-7}$ ph cm$^{-2}$ s$^{-1}$. Here, 
95\% upper limits are shown by vertical arrows. Vertical dashed lines indicate the period for which SED has been made.}
\label{fig2}
\end{figure*}

\begin{figure}
\includegraphics[scale=0.8]{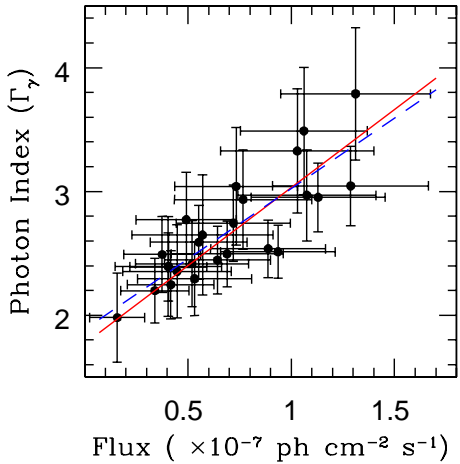}
\caption{$\gamma$-ray photon flux v/s photon-index 
for the source PKS 1502+036. Here, dashed line (blue color) shows the 
unweighted least squares fit and the solid line (red color) shows the weighted
least squares fit.}
\label{fig3}
\end{figure}

{\label{subsec6}}
To investigate the physical characteristics of these sources, we have built optical/UV to $\gamma$-rays SEDs using quasi-simultaneous data. For PKS 1502+036, we have used the data sets between MJD 55800 and 56130, whereas for PKS 2004$-$447, the data sets spanning the period MJD 55680 and 56130 were used. These time periods are shown as vertical dashed lines in Figure \ref{fig1} and \ref{fig2} for PKS 1502+036 and PKS 2004$-$447 respectively. The derived flux values in optical/UV, X-ray and $\gamma$-rays are given in Table \ref{tab2}. Both the SEDs 
were modeled using a simple one zone leptonic emission model described in \citet{2012MNRAS.419.1660S}. In this model the emission region is assumed to be a spherical blob moving down the jet at relativistic speed with bulk Lorentz factor $\Gamma$. The emission region is filled with a broken power law electron distribution with indices $p$ and $q$ before and after the break energy, respectively. The particles lose their energy through synchrotron radiation in a randomly oriented magnetic field ($B$) and inverse Compton radiation. The target photons for the inverse Compton radiation are the synchrotron photons (SSC) and an external photon field which is assumed to be a blackbody for simplicity. Considering the magnetic field to be in equipartition with the relativistic particle distribution, the main parameters governing the SED can be deduced/constrained using the observed information available
in optical/X-ray and $\gamma$-ray energies. The kinetic power of the jet can then be calculated by assuming the inertia of the jet being provided by cold protons whose number density is equal to that of non-thermal electrons. This model does not take into account UV/optical emission from the accretion disk and X-ray emission from the hot corona. For the model fitting we assume a variability time scale of $\sim$1 day, which is typically seen in the $\gamma$-ray lightcurves
of  blazars. 

\begin{figure*}
\hspace{-1.4cm}
\hbox{
\includegraphics[width=8.8cm]{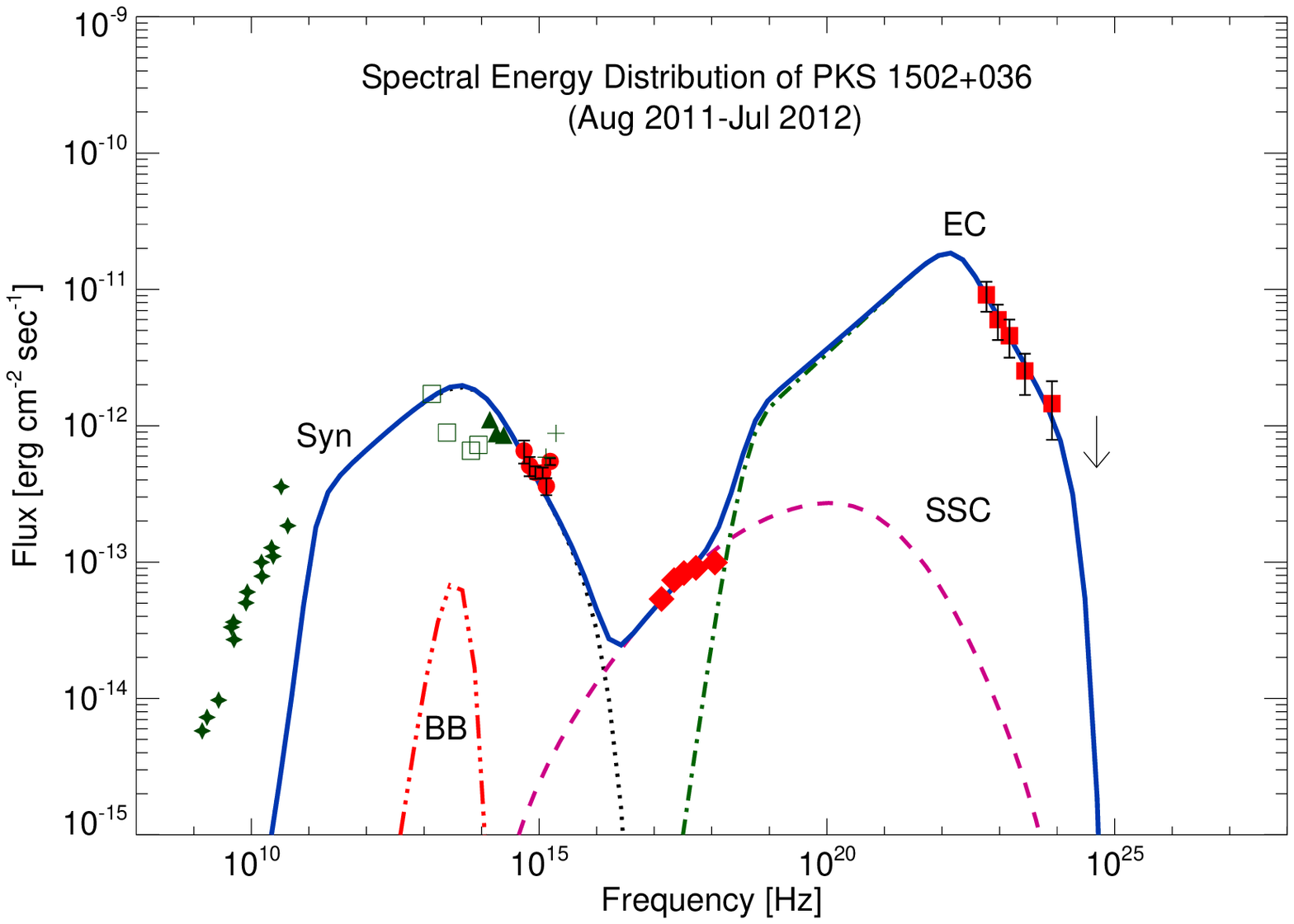}
\includegraphics[width=8.8cm]{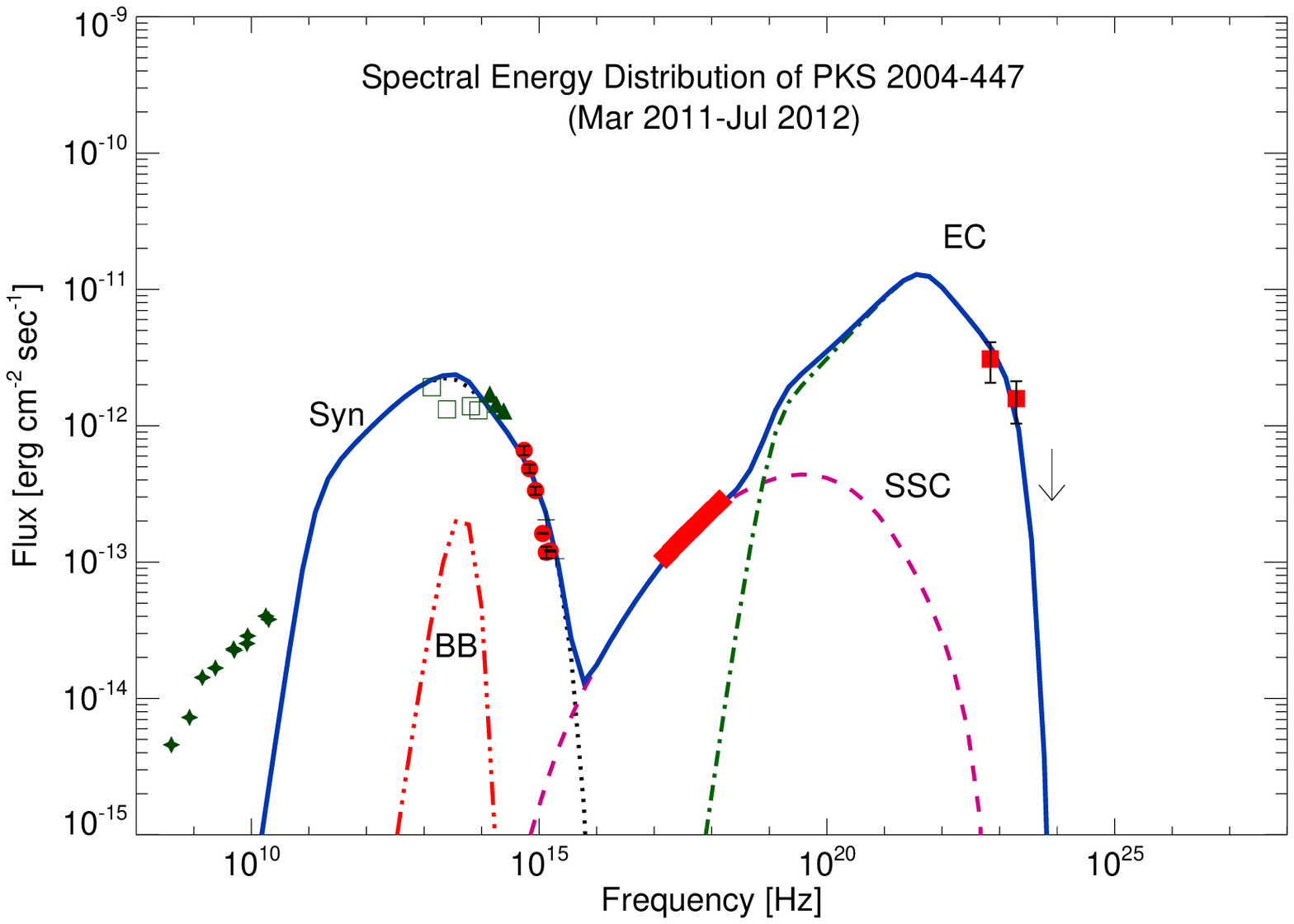}
}
\caption{SEDs of $\gamma$-NLSy1 galaxies PKS 1502+036 and PKS 2004$-$447. Flux due to {\it Fermi}-LAT (filled squares), {\it Swift} XRT (filled diamonds) and UVOT (filled circles) are indicated with red symbols. Vertical arrow shows the upper limit of $\gamma$-ray flux. Archival data are shown in green color: 2MASS (upward 
triangles \citealt{2006AJ....131.1163S}), WISE (open squares;  \citealt{2010AJ....140.1868W}), GALEX (plus sign; \citealt{2005ApJ...619L...1M}) and Radio 
(filled star; NED). Dotted black line shows synchrotron emission. The SSC (pink color) and EC (green color) processes are shown by dashed and dotted-dashed line, respectively. Blue continous line is the sum of all radiative components. Around 10$^{11}$ Hz, synchrotron self absorption is clearly visible.}
\label{fig4}
\end{figure*}

The resultant model fits to the SEDs for PKS 1502+036 and PKS 2004$-$447 are shown in Figure \ref{fig4}. The optical-UV flux can be reproduced by synchrotron emission 
and the X-ray flux by SSC emission. The explanation of the $\gamma$-ray flux as an outcome of EC process demands the temperature of the external photon field to be $605\;K$ 
in case of PKS 1502+036 and $694\;K$ for PKS 2004$-$447. The spectrum of this external photon field black body component with luminosities of 8.19 $\times$ 10$^{43}$ 
erg s$^{-1}$ and 6.26 $\times$ 10$^{43}$ erg s$^{-1}$  for PKS 1502+036 and PKS 2004$-$447 respectively is shown in Figure \ref{fig4}. We compared the 
luminosity of this black body component with the observed luminosity (which is the sum of thermal and non-thermal components) in 2MASS and WISE bands and
as expected they are lower than the observed luminosities. Also the deduced temperatures are consistent with the 
temperature generally observed for the dusty torus (\citealt{2004Natur.429...47J}). 
The archival data from NED is shown with green color in Figure \ref{fig4}. The radio observation may be contaminated by the core emission and emissions from other extended region and hence do not satisfy the model curve. The model fitting parameters are given in Table \ref{tab3}. 
We note that many of the parameters obtained from our modelling of the 
broad band SED of these two sources differ from the paramaters obtained 
for the same sources by \citet{2009ApJ...707L.142A}. This might be partly
due to the quality of the data sets used for SED modelling.
Here we use nearly simultaneous data sets compared to the non-simultaneous
multi-wavelength data used by \citet{2009ApJ...707L.142A}. 

\section{Discussion}{\label{sec5}}
Aligned relativistic jets are invoked to explain the extreme power generated by the blazar class of AGN, with a large fraction of the power being emitted in the $\gamma$-ray band. RL-NLSy1 galaxies also have many observed characteristics similar to blazars, namely, flat/inverted radio spectra, compact core-jet structure on VLBI scales \citep{2011A&A...528L..11G,2012arXiv1205.0402O}, high brightness temperature, superluminal jet components \citep{2006PASJ...58..829D}, INOV (\citealt{2013MNRAS.428.2450P,2010ApJ...715L.113L}) and 
rapid infra-red flux variability (\citealt{2012ApJ...759L..31J}). However, the detection of variable $\gamma$-ray emission in five RL-NLSy1 
galaxies \citep{2009ApJ...699..976A,2009ApJ...707L.142A,2012MNRAS.426..317D,2011MNRAS.413.2365C} clearly shows that NLSy1 galaxies could also
have relativistic jets similar to the blazars hosted in elliptical galaxies. If the NLSy1 galaxies are hosted in spiral galaxies 
we should revisit the  paradigm that jets can only be launched in elliptical galaxies (\citealt{2009arXiv0909.2576M}). The blazar like behavior shown by these $\gamma$-NLSy1 galaxies, powered by smaller mass black holes and accreting close to the Eddington limit, also, poses questions on the nature of them vis-a-vis blazars and their place in the traditional blazar sequence. Thus an understanding of these sources will enable one to probe jet physics at a different mass range compared to blazars. We have presented here a detailed analysis of two such 
RL-NLSy1 galaxies namely PKS 1502+036 and PKS 2004$-$447 using multi-wavelength data and modeling. Our multi-wavelength lightcurves clearly show the objects to be  variable; however, owing to the sparseness of the data, correlations if any on the flux variations between different frequency bands could not be ascertained.

\begin{figure*}
\hspace{-1.0cm}
\hbox{
\includegraphics[width=8.5cm]{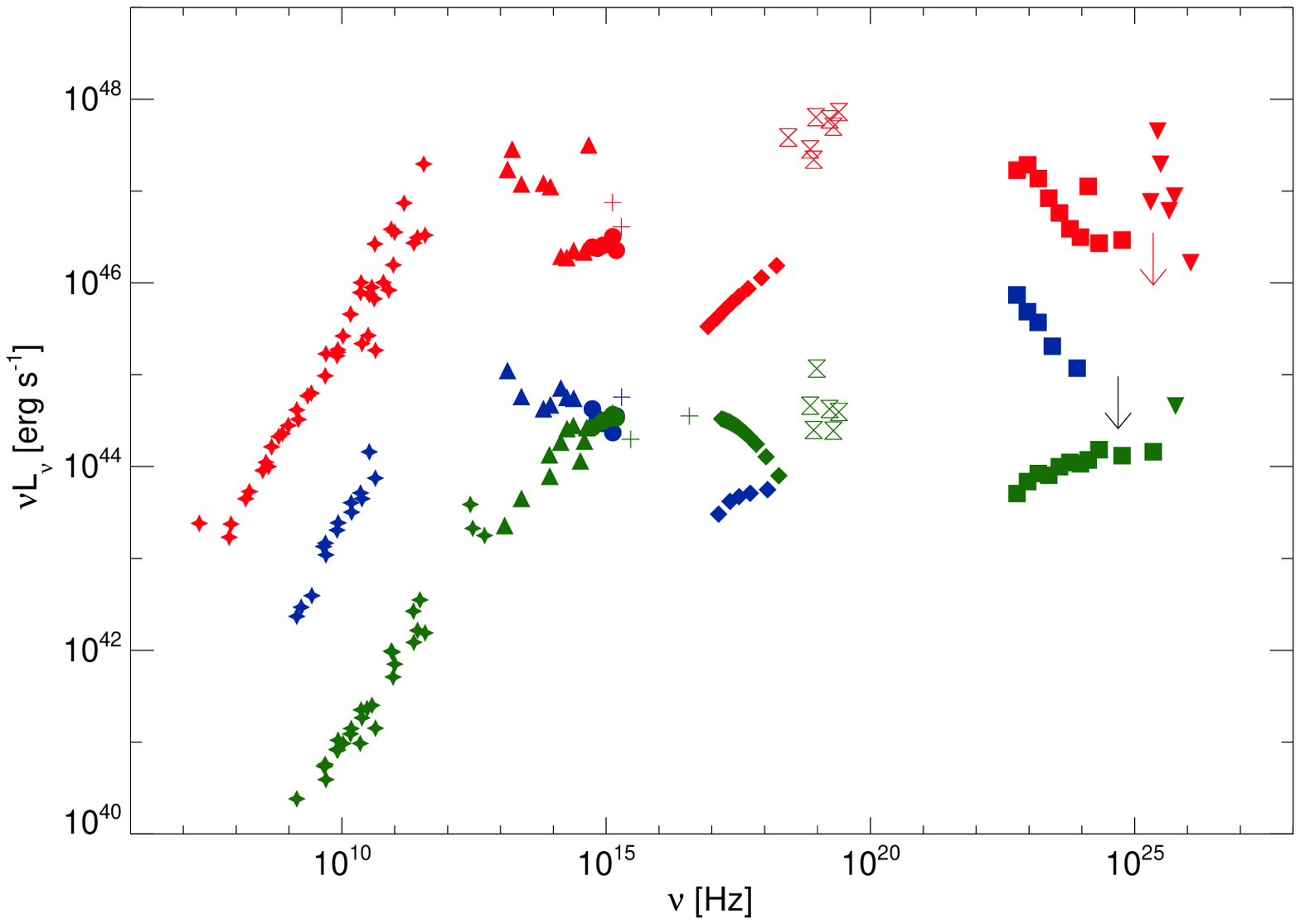}
\includegraphics[width=8.5cm]{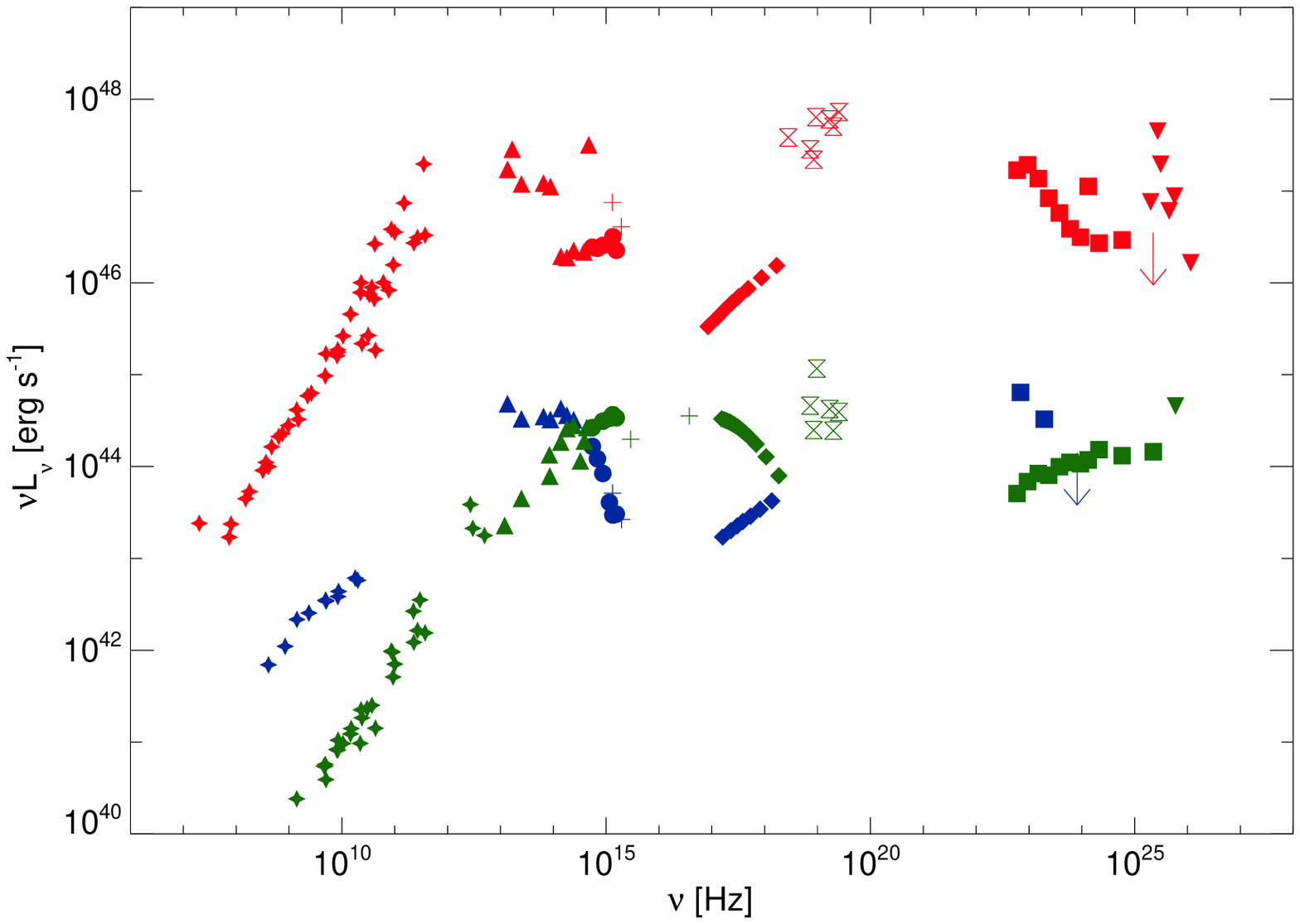}
}
\caption{SEDs of $\gamma$-NLSy1 galaxies PKS 1502+036 (left panel) and 
PKS 2004$-$447 (right panel) along with that of 3C 454.3 and Mrk 421. 
Spectral data points of $\gamma$-NLSy1 galaxies are shown with blue colors. 
Data points of 3C 454.3 (upper side of the plot) and Mrk 421 (lower side of the plot) are shown 
with red and green colors respectively.  Vertical arrows show the upper limit 
of {\it Fermi}-LAT $\gamma$-ray flux.   {\it Fermi}-LAT data is shown with filled squares, while {\it Swift} XRT and UVOT data are shown with filled diamonds and filled circles respectively. Archival radio data is shown with filled stars, whereas IR, UV and hard X-ray data are shown with filled upward triangles,  plus sign and hourglass respectively. Filled downward triangles show the 
MAGIC (\citealt{2009A&A...498...83A}) upper limit data of 3C 454.3 and HAGAR (\citealt{2012A&A...541A.140S}) data of Mrk 421.}
\label{fig5}
\end{figure*}

From the modeling of the SEDs of both PKS 1502+036 and PKS 2004$-$447, using the one zone leptonic model of  \citet{2012MNRAS.419.1660S}, we find that, for these sources, the UV-optical data are well fit with synchrotron emission, the X-ray data are fit with SSC and the $\gamma$-ray points are fit with EC scattering of IR photons from the torus. The power of the jet estimated from the fit parameters is larger for PKS 1502+036 compared to PKS 2004$-$447. To see how the SEDs of these 
two $\gamma$-NLSy1 galaxies match with that of blazars, we also  built 
two other SEDs, one for the well studied FSRQ 3C 454.3 ($z$ = 0.859) and the other for the BL Lac Mrk 421 ($z$ = 0.030).
For both 3C 454.3 and Mrk 421, we utilized data that covered the same period as that used for the two $\gamma$-NLSy1 galaxies studied here.
The data for both 3C 454.3 and
Mrk 421 were analyzed using the procedures outlined in Section \ref{sec3} except
for the XRT analysis of Mrk 421 wherin we have used a log-parabola model (\citealt{2004A&A...413..489M,2004A&A...422..103M}). The results of the spectral analysis for 3C 454.3 and Mrk 421 are given in Table \ref{tab4}. In Figure \ref{fig5}, we show the SEDs of PKS 1502+036 and PKS 2004$-$447 along with the FSRQ 3C 454.3 and BL Lac Mrk 421. In this Figure too, the archival data are from NED and BZCAT\footnote[7]{http://www.asdc.asi.it/bzcat/}. We found that during the 
period considered for spectral analysis, 3C 454.3 was in quiescent state 
while Mrk 421 was in exceptional bright state.

From Figure \ref{fig5}, it is clear that for 3C 454.3, the synchrotron emission peaks in the IR region, whereas the high energy EC emission peaks in the 
MeV$-$GeV range. However, for Mrk 421, the synchrotron emission peaks in 
the UV$-$X-ray spectral band and the high
energy SSC emission peaks in the GeV$-$TeV region. While the optical/UV luminosities of the two $\gamma$-NLSy1 galaxies are comparable to that of Mrk 421, their $\gamma$-ray luminosities are significantly higher than the flaring Mrk 421. This clearly demonstrates that SSC alone cannot account for the high energy emission from these two $\gamma$-NLSy1 galaxies and that EC is needed. 
Such EC model is frequently 
invoked to explain the high energy emission from FSRQs (\citealt{1992A&A...256L..27D,1994ApJ...421..153S,2009ApJ...704...38S}). In the powerful FSRQ, PKS 1510$-$089, based on the observed coincidence of $\gamma$-ray flux to the appearance of new knots, \citet{2010ApJ...710L.126M} hint that the $\gamma$-rays are produced from a region well beyond the radius of the BLR, most likely in the IR torus. Thus, the high energy emission from these $\gamma$-NLSy1 galaxies clearly resemble that of FSRQs and likely due to IC scattering of IR photons from the dusty torus. Another supporting fact for the above mentioned theory is that nominal temperature of molecular torus is about $\sim$ 300 K \citep{2010MNRAS.408.1982L} to $\sim$ 1200 K \citep{2007ApJ...660..117C,2011ApJ...732..116M} with hot emitting region expected to be nearer to the central super massive black holes (SMBHs). The IR torus temperature which we obtained from SED analysis is about $\sim$ 600$-$700 K for both sources, which means that the dissipation region may lie outside the BLR. 
Recently, \citet{2012arXiv1210.5294Z} too have noted that the observed broad band SEDs of four NLSy1 galaxies detected by {\it Fermi} are in close resemblance to FSRQs than BL Lacs. In the SEDs of these NLSy1 galaxies (see Figure \ref{fig5}), it is clear that the low energy bump is in the IR regime, while the high energy bump is in the MeV$-$GeV band. Also, it is clear from Figure \ref{fig5}, that both the low and high energy peaks of these $\gamma$-NLSy1 galaxies 
are intermediate to that of 3C 454.3 and Mrk 421. Thus, both PKS 1502+036 
and PKS 2004$-$447 (and may be other $\gamma$-NLSy1 galaxies) could fit 
very well into the traditional blazar sequence (\citealt{1998MNRAS.299..433F,2010ASPC..427..243F}).

\begin{table*}
\begin{center}
{
\scriptsize
\caption{Summary of SED analysis\label{tab2}}
\begin{tabular}{lcccccccc}
\tableline\tableline
 & & & {\it Fermi}-LAT &  & & & &  \\
\tableline
 Source & Time covered\tablenotemark{a}& Flux\tablenotemark{b} & Photon Index\tablenotemark{c} & Test Statistic\tablenotemark{d} & & & &  \\
 \tableline
 PKS 1502+036 & 55800-56130& 2.1$\pm$0.31 &2.93$\pm$0.19 & 87.17& & & & \\
 PKS 2004-447&55680-56130 &0.71$\pm$0.19 & 2.93$\pm$0.30& 19.17& & & &  \\
 \tableline
 & & & {\it Swift}-XRT  & & & &  \\
 \tableline
 Source& Time covered\tablenotemark{a}& Exp.\tablenotemark{e}& $\Gamma$\tablenotemark{f}& Flux\tablenotemark{g} & Normalization\tablenotemark{h}& Stat.\tablenotemark{i}&  \\
\tableline
 PKS 1502+036& 56042-56103& 14.58&  1.88$\pm$0.22& 2.88$\pm$0.72& 5.28$\pm$0.82& 14.26/21 & \\
 PKS 2004-447& 55696-56130& 45.89 & 1.53$\pm$0.09& 6.26$\pm$0.73& 8.14$\pm$0.57& 36.69/37& \\
 \tableline
 & & & {\it Swift}-UVOT & & & & &  \\
 \tableline
 Source & Time covered\tablenotemark{a}& v\tablenotemark{j}& b\tablenotemark{j}& u\tablenotemark{j}& uvw1\tablenotemark{j}& uvm2\tablenotemark{j}& uvw2\tablenotemark{j}& \\
 \tableline
 PKS 1502+036 & 56042-56130& 6.54$\pm$1.26& 5.08$\pm$0.82& 4.56$\pm$0.48& 4.53$\pm$0.40& 3.61$\pm$0.51& 5.47$\pm$0.32& \\
 PKS 2004-447 & 55696-56130& 6.59$\pm$0.52& 4.84$\pm$0.34& 3.35$\pm$0.23& 1.62$\pm$0.02& 1.18$\pm$0.12& 1.21$\pm$0.02& \\
 \tableline
\end{tabular}
\tablenotetext{1}{Time covered for analysis, in MJD.}
\tablenotetext{2}{Integrated $\gamma$-ray flux in 0.2$-$300 GeV energy range in the units of 10$^{-8}$ ph cm$^{-2}$ s$^{-1}$.}
\tablenotetext{3}{Photon index calculated from $\gamma$-ray analysis.}
\tablenotetext{4}{Significance of detection using likelihood analysis.}
\tablenotetext{5}{Net exposure in kiloseconds.}
\tablenotetext{6}{Photon index of power-law model.}
\tablenotetext{7}{Observed flux in the units of 10$^{-13}$ erg cm$^{-2}$ s$^{-1}$, in 0.3$-$10 keV energy band.}
\tablenotetext{8}{Normalization at 1 keV in 10$^{-5}$ ph cm$^{-2}$ s$^{-1}$ keV$^{-1}$.}
\tablenotetext{9}{Statistical parameters:$\chi^{2}$/dof.}
\tablenotetext{10}{Average flux in {\it Swift} V, B, U, W1, M2 and W2 bands respectively, in units of 10$^{-13}$ erg cm$^{-2}$ s$^{-1}$.}
}
\end{center}
\end{table*}

\begin{table*}
\begin{center}
{
\caption{Summary of model parameters for the SED of PKS 1502+036 and PKS 2004$-$447\label{tab3}}
\begin{tabular}{lccc}
\tableline
Parameter& Symbol& PKS 1502+036& PKS 2004$-$447\\
\tableline
Redshift & $z$& 0.409& 0.240\\
Particle spectral index (low energy)& p& 2.22& 2.10\\
Particle spectral index (high energy) & q& 4.5& 4.00\\
Magnetic field (equipartition)& B& 0.60 G & 0.72 G\\
Bulk Lorentz factor & $\Gamma$& 12 & 9\\
Comoving emission region size& R$\textquotesingle$& 2.86 $\times$ 10$^{16}$ cm & 2.32 $\times$ 10$^{16}$ cm\\
IR dust temperature in AGN frame& {\it T}$^{*}$& 605 K& 694 K\\
IR torus energy density in AGN frame& {\it u}$^{*}$& 1.02 $\times$ 10$^{-3}$ erg cm$^{-3}$& 1.75 $\times$ 10$^{-3}$ erg cm$^{-3}$\\
IR black body luminosity in AGN frame  & L$^{*}$  & 8.19 $\times$ 10$^{43}$ erg s$^{-1}$ &6.26 $\times$ 10$^{43}$ erg s$^{-1}$\\
Particle energy density & {\it U}$_{e}$& 1.43 $\times$ 10$^{-2}$ erg cm$^{-3}$& 2.06 $\times$ 10$^{-2}$ erg cm$^{-3}$\\
Minimum electron Lorentz factor & $\gamma\textquotesingle_{min}$& 30& 60\\
Break electron Lorentz factor& $\gamma\textquotesingle_{b}$& 1870& 1176\\
Maximum electron Lorentz factor & $\gamma\textquotesingle_{max}$& 1.8 $\times $ 10$^{4}$& 7.5 $\times $ 10$^{3}$\\
Jet power & {\it P}$_{j}$& 3.1 $\times$ 10$^{45}$ erg s$^{-1}$&8.5 $\times$ 10$^{44}$ erg s$^{-1}$\\
\tableline
\end{tabular}
}
\end{center}
\end{table*}

\begin{table*}
\hspace{-1.0cm}
\begin{center}
{
\scriptsize
\caption{Summary of the comparison source analysis.\label{tab4}}
\begin{tabular}{lcccccccccc}
\tableline\tableline
 & & & {\it Fermi}-LAT &  & & & & & &\\
\tableline
 Source & Time covered& Flux & Photon index & Test Statistic & & & & & &\\
 \tableline
 3C 454.3 & 55680-56130& 4.9$\pm$0.29 &2.61$\pm$0.06 & 742.73& & & & & &\\
 Mrk 421&55680-56130 &11.4$\pm$0.27 & 1.79$\pm$0.02& 10971.46& & & & & &\\
 \tableline
 & & & {\it Swift}-XRT & & & & & & &\\
 \tableline
 Source& Time covered& Exp.& N$_{H}$& $\alpha$\tablenotemark{a}& $\beta$\tablenotemark{a} &$\Gamma$ & Flux \tablenotemark{b}& Normalization\tablenotemark{c}& Stat.&  \\
\tableline
 3C 454.3& 55729-55924& 17.41& 6.58& $-$& $-$& 1.49$\pm$0.05 &(6.54$\pm$0.38) $\times$ 10$^{-12}$& (8.43$\pm$0.35) $\times$ 10$^{-4}$& 86.77/74 & \\
 Mrk 421& 55621-56078& 112.4 & 1.91& 2.308$\pm$0.001& 0.437$\pm$0.004&$-$ &(3.42$\pm$0.01) $\times$ 10$^{-10}$& (8.96$\pm$0.01) $\times$ 10$^{-2}$& 1177.25/675& \\
 \tableline
 & & & {\it Swift}-UVOT& & & & & &\\
 \tableline
 Source & Time covered& v\tablenotemark{4}& b\tablenotemark{4}& u\tablenotemark{4}& uvw1\tablenotemark{4}& uvm2\tablenotemark{4}& uvw2\tablenotemark{4}& & \\
 \tableline
 3C 454.3 & 55729-55924& 7.13$\pm$0.13& 6.94$\pm$0.10& 7.46$\pm$0.13& 7.63$\pm$0.16& 9.28$\pm$0.16& 6.59$\pm$0.13& & \\
 Mrk 421 & 55621-56078& 132.98$\pm$4.34 & $-$ & 156.31$\pm$8.24 & 167.59$\pm$1.04 & 183.96$\pm$8.48 & 169.82$\pm$9.91 & & \\
 \tableline
\end{tabular}
\tablenotetext{1}{Log-parabola model parameters used in XSPEC}
\tablenotetext{2}{0.3-10 keV flux, in the units of erg cm$^{-2}$ s$^{-1}$}
\tablenotetext{3}{normalization having units as ph cm$^{-2}$ s$^{-1}$ keV$^{-1}$}
\tablenotetext{4}{{\it Swift} V, B, U, W1, M2 and W2 band fluxes in the units of 10$^{-12}$ erg cm$^{-2}$ s$^{-1}$}
\tablecomments{See Table \ref{tab2} for description of remaining parameters.}
}
\end{center}
\end{table*}

Recently, \citet{2011ApJ...735..108C} has reported that the $\gamma$-NLSy1 galaxies occupy low luminosity-low frequency region in synchrotron peak frequency ($\nu_{s}$)-luminosity ({\it L}$_{s}$) plane. This trend can also be very well seen in Figure \ref{fig5}, where {\it L}$_{s}$ of both PKS 1502+036 and PKS 2004$-$447 are similar to that of Mrk 421, whereas, their corresponding peak frequencies are quite different. Both of them have $\nu_{s}$ in the lower frequency region compared to Mrk 421. Again the possible explanation for such low luminosity - low frequency behaviour of both PKS 1502+036 and PKS 2004$-$447 could be due to 
their low black hole masses \citep{2011ApJ...735..108C,2008MNRAS.387.1669G}.

\begin{figure}
\hspace{-1.0cm}
\centering
\includegraphics[width=9cm,height=9cm]{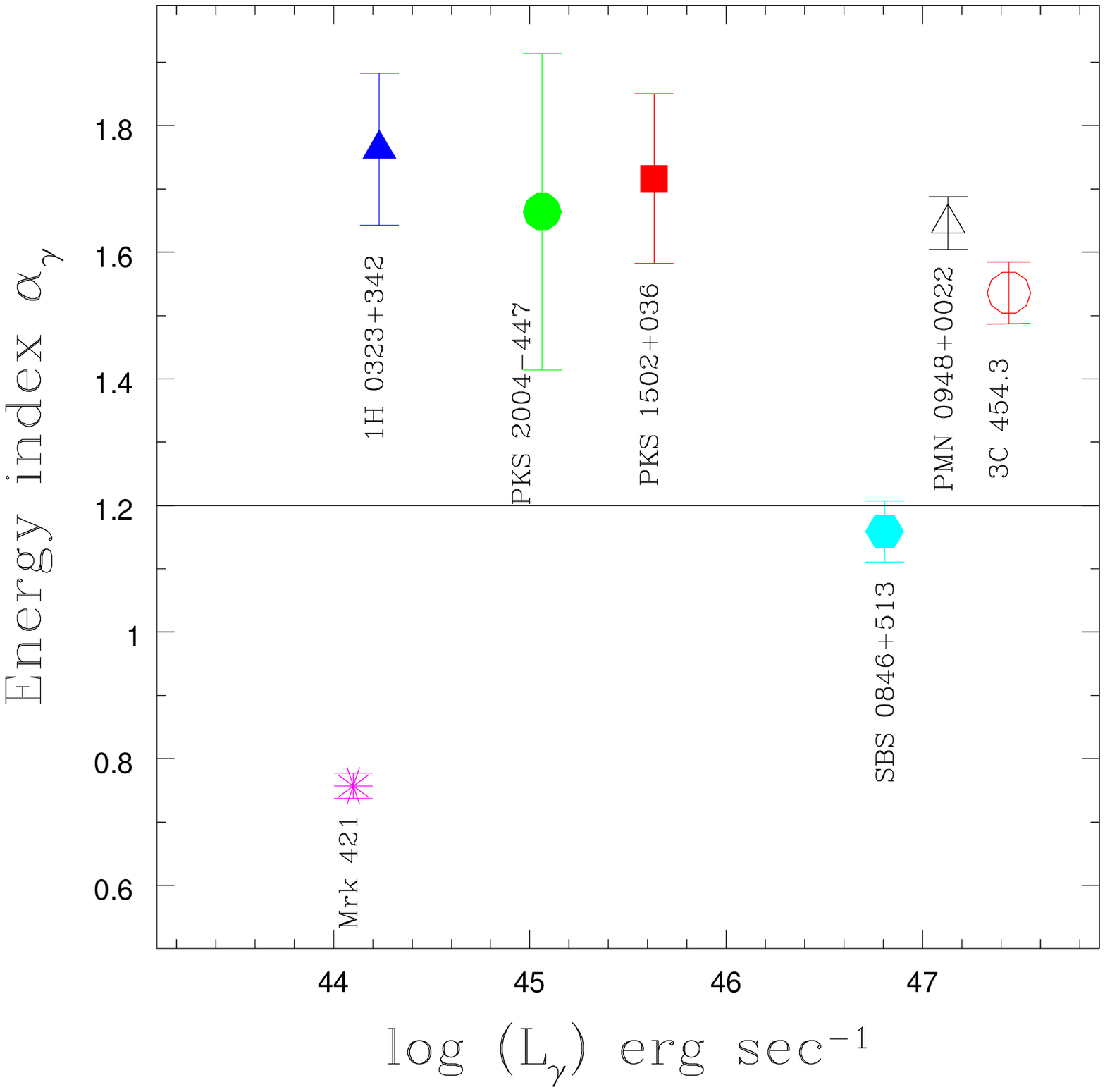}
\caption{The spectral index ($\alpha_{\gamma}$) against  
the Luminosity ({\it L}$_{\gamma}$) in the energy band 0.1-10 GeV. 
The horizontal grey line marks $\alpha_{\gamma}$ = 1.2.}
\label{fig6}
\end{figure}

In Figure \ref{fig6}, we have shown the position of the two NLSy1 galaxies studied here, in the $\gamma$-ray spectral index v/s K-corrected $\gamma$-ray luminosity plane along with the FSRQ 3C 454.3 and the BL Lac Mrk 421. Also shown in Figure \ref{fig6}, are the other three known $\gamma$-ray emitting NLSy1 galaxies namely 1H 0323+342 ($z$ = 0.063), SBS 0846+513 ($z$ = 0.584) and 
PMN J0948+0022 ($z$ = 0.584). The K-corrected $\gamma$-ray luminosities were evaluated following \citet{2009MNRAS.396L.105G}. Considering only the sources studied here, namely, PKS 1502+036 and PKS 2004$-$447, we find that their $\gamma$-ray luminosities are intermediate to the FSRQ 3C 454.3 and BL Lac Mrk 421 and, they have 
steep $\gamma$-ray spectral index similar to FSRQs (\citealt{2009MNRAS.396L.105G}). However, considering all the {\it Fermi} detected NLSy1 galaxies till now, 
we found that they have a wide range of $\gamma$-ray luminosities overlapping the region occupied by BL Lacs and FSRQs (\citealt{2009MNRAS.396L.105G,2011MNRAS.414.2674G}), however, the steep $\gamma$-ray spectral index displayed by most of them, leads us to argue that in terms of the $\gamma$-ray spectral index behavior, {\it Fermi} detected NLSy1 galaxies are similar to FSRQs.

Further, as discussed by \citet{2009IJMPD..18.1483C}, FSRQs are more Compton-dominated (Compton-dominance is the ratio between Compton and synchrotron luminosities, CD $\equiv$ {\it L$_{C}/L_{s}$}) than BL Lacs and on average show a sequence of values from $\sim$100 in FSRQs to $\lesssim$ 1 in high energy peaked BL Lac objects (HBLs). If we calculate CD for our sources, it comes $\sim$ 10 for both of them. Finally, X-ray spectra of low-energy peaked BL Lac objects (LBL) and FSRQs are believed to be dominated by IC emission of low-energy electrons and thus a flat spectral index ($\alpha_{x}$ $<$ 1) is expected. As we move from LBL to HBL sources, the peak shifts toward higher energies and tail of the synchrotron emission starts to dominate  and a steep spectrum ($\alpha_{x}$ $>$ 1) is expected.  If we calculate the X-ray spectral indices for our sources (see Table \ref{tab2} and \ref{tab4}), 
we found that both PKS 1502+036 and PKS 2004$-$447 are having flat spectral indices (0.88 and 0.53, respectively), which are intermediate to 3C 454.3 (0.49) and Mrk 421 (1.31).

Thus, from an analysis of the presently available multi-wavelength data set of PKS 1502+036 and PKS 2004$-$447, we find that many properties of them are intermediate to the FSRQ 3C 454.3 and the BL Lac Mrk 421. Further, our analysis also 
 suggests that these sources may be low black hole mass counterparts of powerful FSRQs. However, we caution that as of now there are no unambiguous observational evidence for these two sources in particular and NLSy1 galaxies in general, to be hosted by spiral galaxies having low mass black holes. However, going by the present widely considered notion that NLSy1 galaxies are powered by low mass black holes in spiral galaxies, the current observational evidence for the presence of aligned jets in the NLSy1 galaxies detected by {\it Fermi} goes against the prevailing idea of jets being launched only by elliptical galaxies. A clear picture on the nature of these sources will emerge when more observations become available in the future. 

\section{Conclusions}{\label{sec6}}
In this work, we have analyzed the two intermediate redshifted $\gamma$-NLSy1 galaxies using all available data covering IR, optical, UV, X and $\gamma$-ray regimes to understand their physical characteristics. We also compare the results obtained, with two other well studied sources namely the FSRQ 3C 454.3 and the BL Lac Mrk 421. We arrive at the following conclusions.

\begin{enumerate}
\item The broad-band SEDs of PKS 1502+036 and PKS 2004$-$447 resemble more, although less powerful, to the SEDs of the FSRQ class of AGN. A possible
interpretation is that they could be the low black hole mass counterparts of powerful 
FSRQs. However, the potential connection between NLSy1 galaxies and FSRQs needs further detailed investigation.
\item The Compton dominance and X-ray spectral index of PKS 1502+036 and PKS 2004$-$447 are between the values found for 3C 454.3 and Mrk 421. In the $\gamma$-ray luminosity $-$ spectral index plane, these sources have luminosities intermediate to that of 3C 454.3 and Mrk 421, however, they have steep $\gamma$-ray spectral index similar to the FSRQ 3C 454.3. On inclusion of the other three $\gamma$-ray emitting NLSy1 galaxies (see Figure \ref{fig6}), we found that $\gamma$-NLSy1 galaxies may have a wider range of $\gamma$-ray luminosities between FSRQs and BL Lacs, however, they have $\gamma$-ray spectral index more like FSRQs. The derived spectral properties of the two sources studied here are, in general, similar to blazars and intermediate between the FSRQ 3C 454.3 and BL Lac Mrk 421. They thus, could fit well into the traditional blazar sequence.
\item In the $\gamma$-ray band, spectral softening when brightening trend is clearly seen for PKS 1502+036. Similar trend is also seen in PKS 2004$-$447 between the available two epochs of observations. This is against the harder when brighter trend seen in the FSRQ 3C 454.3.
\end{enumerate}

\acknowledgments
This research has made use of the data obtained from the High Energy Astrophysics Science Archive Research Center (HEASARC) provided by the NASA's Goddard Space Flight Center. This research has made use of the NASA/IPAC Extragalactic Database (NED) which is operated by the Jet Propulsion Laboratory, California Institute of Technology, under contract with the National Aeronautics and Space Administration. This research has made use of the XRT Data Analysis Software (XRTDAS) developed under the responsibility of the ASI Science Data Center (ASDC), Italy. 



{\it Facilities:} \facility{Fermi}, \facility{Swift}.

\bibliographystyle{apj}
\bibliography{my}

\end{document}